\documentclass[journal]{IEEEtran}
\usepackage{cite}
\usepackage{amsmath,amssymb,amsfonts}
\usepackage{algorithmic}
\usepackage{graphicx}
\usepackage{textcomp}

\usepackage{multirow}
\usepackage{subfig}
\usepackage{soul}
\usepackage{color}
\usepackage{bm}
\usepackage{hyperref}

\def\BibTeX{{\rm B\kern-.05em{\sc i\kern-.025em b}\kern-.08em
    T\kern-.1667em\lower.7ex\hbox{E}\kern-.125emX}}
\begin{document}

\title{Dual-domain Cascade of U-nets for Multi-channel Magnetic Resonance Image Reconstruction}

\author{Roberto Souza, PhD, Mariana Bento, PhD, Nikita Nogovitsyn, MSc, MD, Kevin J. Chung, BSc, R. Marc Lebel, PhD, and~Richard Frayne, PhD 
\thanks{R Souza, M Bento, KJ Chung and R Frayne are with the Departments of Radiology and Clinical Neuroscience, Hotchkiss Brain Institute, University of Calgary, Calgary, AB, T2N 1N4 Canada, and the Seaman Family MR Research Centre, Foothills Medical Centre, Alberta Health Services, Calgary, Alberta, T2N 2T9. N Nogovitsyn is with the Department of Psychiatry, Hotchkiss Brain Institute, University of Calgary, AB, T2N 1N4 Canada, and the Mathison Centre for Mental Health Research and Education, University of Calgary, Calgary, AB, T2N 1N4 Canada. RM Lebel is with General Electric Healthcare, Calgary, AB, T2E 7E2 Canada.
 E-mails: {roberto.medeirosdeso@ucalgary.ca, mariana.pinheirobent@ucalgary.ca, n.nogovitsyn@ucalgary.ca, kjy.chung@gmail.com, Marc.Lebel@ge.com, rfrayne@ucalgary.ca}} 
\thanks{arXiv preprint, November 2019.}
}

\markboth{arXiv}
{Shell \MakeLowercase{\textit{et al.}}: Bare Demo of IEEEtran.cls for IEEE Journals}

\maketitle

\begin{abstract}
The U-net is a deep-learning network model that has been used to solve a number of inverse problems. In this work, the concatenation of two-element U-nets, termed the W-net, operating in k-space (K) and image (I) domains, were evaluated for multi-channel magnetic resonance (MR) image reconstruction. The two element network combinations were evaluated for the four possible image-k-space domain configurations: a) W-net II, b) W-net KK, c) W-net IK, and d) W-net KI were evaluated. Selected promising four element networks (WW-nets) were also examined. Two configurations of each network were compared: 1) Each coil channel processed independently, and 2) all channels processed simultaneously. One hundred and eleven volumetric, T1-weighted, 12-channel coil k-space datasets were used in the experiments. Normalized root mean squared error, peak signal to noise ratio, visual information fidelity  and visual inspection were used to assess the reconstructed images against the fully sampled reference images. Our results indicated that networks that operate solely in the image domain are better suited when processing individual channels of multi-channel data independently. Dual domain methods are more advantageous when simultaneously reconstructing all channels of multi-channel data. Also, the appropriate cascade of U-nets compared favorably ($p < 0.01$) to the previously published, state-of-the-art Deep Cascade model in in three out of four experiments.
\end{abstract}

\begin{IEEEkeywords}
Magnetic resonance imaging, compressed sensing, multi-channel (coil), image reconstruction, inverse problems, brain, machine learning
\end{IEEEkeywords}

\maketitle

\IEEEPARstart{M}{agnetic resonance}  (MR) imaging is a sensitive diagnostic modality that allows specific, high-quality investigation of structure and function of the brain and body. One major drawback is the overall acquisition time to complete an MR imaging protocol, which can easily exceed 30 minutes per patient \cite{zbontar2018fastmri}. Lengthy MR examination times are costly ($\sim$300 USD or more per examination); increase susceptibility to patient motion artifacts, which negatively impact image quality; further reduce patient throughput and contribute to repeated studies. Speed in clinical MR imaging is essential \cite{runge2017speed}. Parallel imaging (PI)\cite{deshmane2012parallel} and compressed sensing (CS)\cite{RN247,RN280} are two proven approaches that allow MR examinations to be shortened. PI techniques, such as Generalized Autocalibrating Partially Parallel Acquisition (GRAPPA)\cite{grappa}, which operates in the spatial-frequency domain (known as k-space), and Sensitivity Encoding for fast MR imaging (SENSE)\cite{sense}, which works in the image domain, are currently both used clinically. CS techniques are more recent and leverage properties of image sparsity to sample k-space at rates lower than the Nyquist sampling theorem and consequently speed-up MR acquisitions. PI and CS can lead to reduction in image acquisition times while preserving diagnostic imaging quality \cite{yang2016sparse}. Both require the implementation of more sophisticated reconstruction algorithms to convert the acquired raw MR signal, k-space, into an appropriate image domain representation.

PI has been clinically adopted but still at conservative speed-up  rates. CS reconstruction methods face challenges in clinical adoption and/or application due to a series of factors, such as the difficulty to validate CS approaches due to limited understanding of artifacts that can arise from sparse reconstruction techniques \cite{yang2016sparse}. Also, while shortening acquisition times, traditional PI and CS approaches can considerably lengthen reconstruction times to a point that reconstructions need to be computed offline, sometimes completing only after the patient has left the scanner. This is a challenging problem, in particular for traditional CS methods, where reconstruction is iterative and often requires heuristic tuning of parameters.   Using deep learning \cite{RN250} methods to reconstruct both fully sampled and especially, undersampled data, is an active research area that can potentially reduce MR costs by reducing acquisition and reconstruction times, and increase patient scanning efficiency over traditional PI and CS methods. 

Recent studies have begun to demonstrate that MR image reconstruction algorithms based on deep learning methods could provide even greater flexibility without compromising image quality \cite{RN254,qin2019convolutional,RN305,RN306,xiang2018deep,RN307,zhang2018multi,RN289,mardani2019deep}. Nevertheless, many of these investigations have been limited to single-channel (SC) coil raw data or synthetic data, often generated by taking the Fourier transform (FT) of magnitude MR images resulting in a k-space with Hermitian symmetry \cite{mardani2019deep,eo2018kiki,RN306,RN307,RN305}. These scenarios are less realistic because modern scanners use multi-channel (MC) coils and raw k-space data is not symmetric in practice. In addition, most previous studies have focused on single-domain approaches, operating in either the image or k-space domains, but less frequently in both domains. Due to these and other factors, MR image reconstruction is considered a frontier for machine learning \cite{RN310}.

In this work, we examined concatenations of U-nets \cite{RN196}, which we term W-nets, across different domain configurations for reconstructing sparsely sampled MC MR data that otherwise would typically undergo an iterative CS reconstruction. We investigated two networks configurations for MC reconstruction (Supplementary Figure 1): 1) \textbf{SC configuration} where each channel of the MC data is processed independently; 2) \textbf{MC configuration} where the MC data is processed simultaneously and the model outputs one image reconstruction per channel. In both models, the  final image reconstruction was obtained by combining the individual channel images with square root sum of squares \cite{larsson2003snr}, which we  referred simply as sum of squares in this work. More specifically, we explored the impact of these model configurations and two different acceleration factors, denoted by \emph{R}, for CS MR reconstruction. The best W-net configuration was also investigated in the context of a deeper network cascade, \textit{i.e.}, WW-net. We compared  W- and WW-nets against the recently published state-of-the-art flat unrolled cascade model (Deep Cascade, \cite{RN306}).

Our hypothesis was that hybrid approaches perform best when using the MC configuration because unlike the SC configuration, they implicitly leverage correlations across channels in a fashion similar to parallel imaging. Also better performance was expected with the W-net because information was processed in both k-space and image domains without needing to learn the domain transform. We expect that the results from the MC configuration would be superior to the SC configuration, because the MC configuration looked at all channels simultaneously. Finally, we expected that cascades of U-nets could potentially outperform flat unrolled cascades. The data used in this work was made publicly available for benchmarking purposes.


\section{Brief Literature Review}

The idea of applying machine learning for MR reconstruction is not new. Nearly  thirty years ago, for example, neural networks were investigated in the context of minimizing Gibbs artifacts resulting from k-space truncation \cite{yan1993data,hui1995mri,hui1995comments}. More recently, hardware and software advancements have allowed for training of advanced models and by 2016, the first deep learning models were being investigated \cite{sun2016deep,wang2016accelerating}. Since then, the deep-learning-based MR reconstruction field has grown rapidly. Several deep-learning models have been proposed for MR CS reconstruction, however most have been validated using private datasets and in a SC acquisition setting. In late 2018, the fastMRI initiative \cite{zbontar2018fastmri} made SC and MC knee raw MR data available for benchmarking purposes. The \emph{Calgary-Campinas} initiative \cite{RN136} has also added publicly available SC brain MR raw data. With this report, we provide access to MC data. 

\begin{figure}[!h]
\centering
\includegraphics[width=0.495\textwidth]{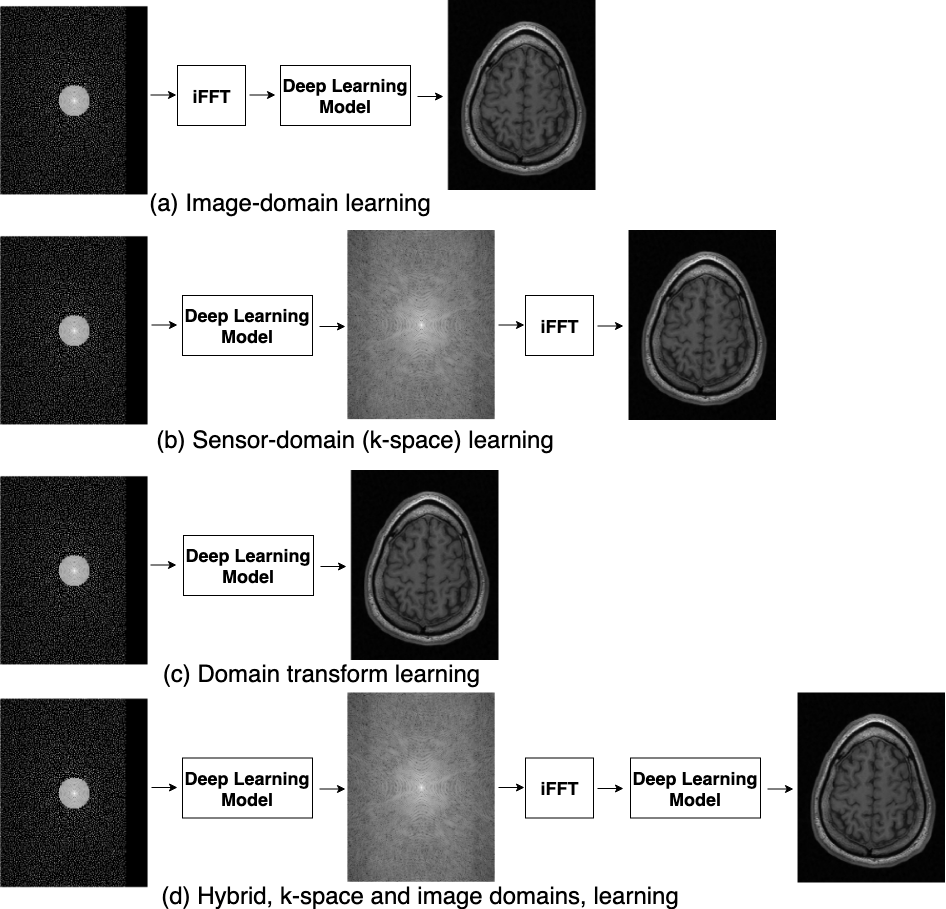}
\caption{Groups of deep learning techniques proposed for magnetic resonance image reconstruction: (a) image domain learning, (b) sensor-domain (k-space) learning, (c) domain transform learning, and (d) hybrid sensor- and image-domain learning. See text for technique descriptions. iFFT = inverse Fast Fourier Transform.}
\label{rec_groups}
\end{figure}

Deep-learning-based MR reconstruction models can be categorized into four groups (Figure \ref{rec_groups}):
\begin{enumerate}
    \item \textbf{Image domain learning} uses an image obtained by inverse Fourier transforming the zero-filled k-space as a starting point. It uses a deep learning model that operates solely in the image domain;
    \item \textbf{Sensor domain (k-space) learning} uses a model operating solely in the acquisition domain, which is the spatial-frequency domain in the case of MR imaging. It tries to estimate the missing k-space samples followed by applying the inverse FT to reconstruct the final image;
     \item \textbf{Domain transform learning} methods try to learn the appropriate transform directly from the sparsely sampled k-space data in order to generate alias free image-domain reconstructions; and
    \item \textbf{Hybrid (sensor and image domains) learning} comprises blocks that process the data in both sensor (k-space) and image domains. These blocks are connected through the appropriate FT (\textit{i.e.}, direct or inverse).
   
\end{enumerate}
The majority of techniques proposed to date are image domain learning methods (see Table \ref{techniques_classification}).

\begin{table}[!h]
\caption{Literature summary classifying key magnetic resonance imaging reconstruction methods into four groups: 1) image domain learning, 2) sensor domain learning, 3)  domain transform learning, and 4) hybrid domain learning.}
\label{techniques_classification}
\centering
\begin{tabular}{c|c}
\hline
 \textbf{Model} &  \textbf{Reference} \\ \hline
 \textbf{Image domain learning} & \cite{mardani2019deep,Schlemper2018StochasticDC,dedmari2018complex,qin2019convolutional,semantic_interpretability,xiang2018deep,RN306,RN307,RN305,RN253,wang2016accelerating,kwon2017parallel,hammernik2018learning,han2018deep,zengK2019}\\ \hline
 \textbf{Sensor domain learning} & \cite{zhang_micccai_2018,akccakaya2019scan,kim2019loraki}\\ \hline
  \textbf{Domain transform learning} & \cite{RN289,RN323,dautomap} \\ \hline
 \textbf{Hybrid domain learning} & \cite{souza19a,wang2018,eo2018kiki,souza2018hybrid}\\ \hline
  \end{tabular}
\end{table}


\subsection{Image Domain Learning}
The seminal work of Jin \textit{et al.} \cite{RN254} proposed to use a direct inversion, which would be the zero-filled k-space inverse FT in the case of MR reconstruction, followed by a residual U-net to solve normal-convolutional inverse problems. The residual connection learns the difference between input and output to mitigate the vanishing gradient problem that can potentially disturb the network training process. Jin \textit{et al.} tested their model by reconstructing x-ray computed tomography in synthetic phantoms and real sinograms, but their model is directly extendable to MR reconstruction. In another study, Lee \textit{et al.}\cite{RN253} compared residual against non-residual U-nets to reconstruct MC MR data. Their results clearly indicated the advantage of using the residual connection, which have subsequently been incorporated in the majority of recently proposed models (\textit{cf.}, \cite{RN307,RN305,RN306,qin2019convolutional,eo2018kiki,mardani2019deep}). 

The model proposed by Schlemper \textit{et al.} \cite{RN306} consists of a flat unrolled deep cascade of CNNs interleaved with data consistency (DC) operations. DC replaces the network k-space signal estimates by measurements obtained in the sampling process. For dynamic MR reconstruction, their model also included data sharing layers. Seitzer \textit{et al.} \cite{semantic_interpretability} built upon \cite{RN306} by adding a visual refinement network, which in their case is a residual U-net, that was trained independently using the result of the flat unrolled deep cascade as its input. Their results showed improvement in terms of semantic interpretability and mean opinion scores\cite{semantic_interpretability}, but the flat unrolled cascade was still better in terms of peak signal-to-noise ratio (pSNR). In a subsequent work, Schlemper \textit{et al.} \cite{Schlemper2018StochasticDC} added dilated convolutions and a stochastic component to their originally proposed model \cite{RN306}.  The dilated convolutions were used to efficiently increase the network receptive field, while the stochastic component consisted of dropping subnetworks of the cascade with a given probability. These authors claimed that their stochastic component accelerated learning because the network became shorter and each subnetwork could see different levels of residual noise, which made the model more robust.

 The cascaded deep learning models consist basically of a stack of convolutional layers. However, 
when the number of convolutional layers increase, their influence on subsequent layers decrease.
It makes the training process more difficult.
In order to overcome this problem, Zeng \textit{et al.}\cite{zengK2019} proposed a very deep densely connected network that combines sub-networks connecting them by dense connections. Each sub-network generates a reconstructed MR image using the information of the previous sub-network. The sub-networks are composed from convolutional and DC layers. The dense connections are expected to help with the vanishing gradient  problem and the training process, improving the overall reconstruction performance of the network. 

A commonly perceived problem with CS MR reconstruction techniques is the loss of high-frequency information, which happens due to two factors: 1) Most k-space sampling schemes favor sampling the low-frequencies more densely, \textit{i.e.}, high frequencies are less densely sampled; and 2) commonly used network loss functions, such as $L_2$ norm, tend to give smooth reconstructions. Adversarial models try to mitigate this problem by including a new term in the reconstruction model (generator)  cost function based on the capacity of a properly trained classifier (discriminator) to distinguish between a fully sampled inverse FT reconstruction and a CS accelerated MR reconstruction. Yang \textit{et al.} \cite{RN307} proposed a deep de-aliasing generative adversarial network (DAGAN) that used a residual U-net as generator with a loss function composed of four different components: an image domain loss, a frequency domain loss, a perceptual loss, and an adversarial loss. Quan \textit{et al.} \cite{RN305} proposed an adversarial model with a cyclic loss \cite{RN315}. Their method consisted of a reconstruction network cascaded with a refinement network governed by a cyclic loss component that tried to enforce that the mapping between input (sampled k-space) and output (reconstructed image) was a bijection, \textit{i.e.} invertible. Mardani \textit{et al.} \cite{mardani2019deep} proposed a generative adversarial network for compressed sensing (GANCS) that tried to model the low dimensional manifold of high-quality MR images by leveraging a mixture of least-squares generative adversarial network and a pixel-wise $L_1/L_2$  cost. 

The work of Dedmari \textit{et al.}\cite{dedmari2018complex}, to the best of our knowledge, is the only study that implemented a complex-valued fully convolutional neural network (CNN) for MR reconstruction. The main advantage of their work was that they took full advantage of the complex number arithmetic as opposed to the other techniques that represent complex-values by splitting real and imaginary components into separate image channels. Unfortunately complex-valued neural networks implementations are still in their infancy, and  deep learning frameworks do not provide support for defining complex networks, which is the reason we did not use them in this work.

We would like to emphasize that even though some image domain learning models described have DC blocks or a frequency-domain term in the network training loss function, the learning portion of these models happened in image domain. Therefore, we did not classify these models as hybrid.

\subsection{Sensor Domain (K-space) Learning}
The work of Zhang \textit{et al.} \cite{zhang2018multi} followed the trend of using adversarial models. The authors proposed a MC generative adversarial network for MR reconstruction in the k-space domain. They tested their approach on 8-channel data using coherent sampling. Their network output estimated the fully sampled k-spaces for each channel. Their final reconstruction was obtained by taking the channel-wise inverse FT and combining the channels through sum of squares \cite{larsson2003snr}. Ak\'{c}akaya \textit{et al.} \cite{akccakaya2019scan} proposed a scan-specific model for k-space interpolation that was trained on the autocalibration signal. Their model outperformed GRAPPA especially for acceleration factors $R>4$. Kim \textit{et al.} \cite{kim2019loraki} proposed a similar approach, but using a recurrent neural network model. In their experiments they outperformed the model proposed in \cite{akccakaya2019scan}.

\subsection{Domain Transform Learning} 
Zhu \textit{et al.} \cite{RN289} proposed to learn the manifold of the transform that connected the sampled k-space and image domains. Their technique is called automated transform by manifold approximation (AUTOMAP). Their model had a quadratic parameters complexity, which did not allow them to train their model due to hardware limitations for images of dimensions greater than $128\times128$ pixels. Subsequent work by Schlemper \textit{et al.} \cite{dautomap} proposed to decompose AUTOMAP (d-AUTOMAP). Instead of learning  a two-dimensional transform, they decomposed it into two one-dimensional transforms, which made their model parameter complexity linear. In their comparison, d-AUTOMAP outperformed AUTOMAP. A somewhat similar approach that looks into the translation of one-dimensional inverse FT of k-space to an image  was investigated in \cite{RN323}. The authors only compared their proposal against traditional CS reconstruction models and demonstrated superior results. 

\subsection{Hybrid Learning}
Hybrid models leverage information as presented in k-space and image domains without trying to learn the domain transform, making the parameter complexity more manageable. A previous study \cite{souza2018hybrid} proposed a hybrid model, which consisted of a k-space U-net connected to an image domain U-net through the inverse FT. Their model was trained end-to-end. However, the model did not have DC steps, and was assessed only on single-coil data. 
Eo \textit{et al.} \cite{eo2018kiki} developed a dual-domain model named KIKI-net that cascaded k-space domain networks with image domain networks interleaved by DC layers and the appropriate domain transform. A similar approach has also been used for computed tomography reconstruction \cite{adler2018learned}. A further investigation of KIKI-net \cite{souza19a} looked at other possible domain configurations for the sub-networks in the cascade and their results indicated that starting the cascade with an image domain sub-network may be advantageous.

\section{Materials and Methods}


\subsection{Dataset}

One hundred and eleven volumetric T1-weighted partially Fourier-encoded hybrid datasets were consecutively acquired as part of the ongoing Calgary Normative Study \cite{tsang2017white}. Data were acquired on a clinical MR scanner (Discovery MR750; General Electric Healthcare, Waukesha, WI) with a 12-channel coil. A three-dimensional, T1-weighted, gradient-recalled echo, sagittal acquisition was employed on presumed healthy subjects (age: $38.7$ years $\pm$ $17.4$ years [mean $\pm$ standard deviation]; range: $20$ years to $80$ years). Acquisition parameters were TR/TE/TI = $6.3$ ms/ $2.6$ ms/ $650$ ms (92 scans) and TR/TE/TI = $7.4$ ms/ $3.1$ ms/ $400$ ms (19 scans), with $170$ to $180$ contiguous $1.0$-mm slices and a field of view of $256$ mm $\times$ $218$ mm. The acquisition matrix size for each channel was $N_{x} \times N_{y} \times N_{z} = 256\times218\times [145,160]$.
In the slice-encoded direction ($k_z$), data were partially collected up to $N_z=[145,160]$ and then zero filled to $N_z=[170,180]$. The scanner automatically applied the inverse FT, using the fast Fourier transform (FFT) algorithms, to the $k_x-k_y-k_z$ k-space data in the frequency-encoded direction, so a hybrid $x-k_y-k_z$ dataset was saved. K-space undersampling was then performed retrospectively in two directions (corresponding to the phase encoding, $k_y$, and slice encoding, $k_z$, directions). Note that the reconstruction problem is effectively a two-dimensional problem  (\textit{i.e.}, in the $k_y-k_z$ plane). The partial Fourier data were reconstructed by taking the channel-wise iFFT of the collected k-spaces and combining the outputs through the conventional sum of squares algorithm that has been shown to be optimal in terms of signal-to-noise ratio for reconstruction of MC MR \cite{larsson2003snr}.  The reconstructed spatial resolution was $1$ mm$^3$.%




The acquired data were used to train, validate and test the proposed SC and MC deep learning reconstruction models. The raw dataset used in this work is publicly available for benchmark purposes as part of the Calgary-Campinas dataset \cite{RN136} (https://sites.google.com/view/calgary-campinas-dataset /home).

\subsection{Cascade of U-net Models}
Let $\mathbf{x} \in \mathbf{C}^{N_y\times N_z\times N_c}$ represent $N_c$ fully-sampled k-spaces, one for each coil channel, of sizes $N_y\times N_z$ pixels. The fully sampled reconstruction $\mathbf{y} \in \mathbf{C}^{N_y\times N_z\times N_c}$ is given by:
\begin{equation}
 \mathbf{y} = \mathcal{F}^{-1}[\mathbf{x}],   
\end{equation}
where $\mathcal{F}$ is the  two-dimensional  FT operator applied across each channel component of the multi-dimensional array. The input for our model is the undersampled and zero-filled set of measurements $\mathbf{x}_u \in \mathbf{C}^{N_y\times N_z\times N_c}$ that can be conveniently defined by:
\begin{equation}
\mathbf{x}_u=F_u\odot \mathbf{x},    
\end{equation}
where  $\odot$ is the element-wise multiplication and $F_u \in \mathbf{R}^{N_y\times N_z\times N_c}$ represents the sampling function defined by:
\begin{equation}
F_u[k_y,k_z,c] = 
\begin{cases}
    1,& \text{if } (k_y,k_z) \in \Omega\\
    0,              & \text{otherwise}
\end{cases}.    
\end{equation}
$\Omega$ is the set of k-space positions sampled. Our models consist of cascading U-nets ($f_{unet}$) where each U-net block operates either on k-space or image domains. The k-space domain U-net ($f_k$):
\begin{equation}
\label{fk}
f_k = f_{unet}(\mathbf{x}_{in})\odot(1-F_u) + \mathbf{x}_u,
\end{equation}
and the image domain U-net ($f_i$):
\begin{equation}
\label{fi}
f_i = \mathcal{F}(f_{unet}(\mathcal{F}^{-1}(\mathbf{x}_{in})))\odot(1-F_u) + \mathbf{x}_u.    
\end{equation}
In these equations, $\mathbf{x}_{in}$ represents a generic input in k-space domain . The right hand side of Equations \ref{fk} and \ref{fi} enforce DC for the k-space positions measured during the sampling process. This DC implementation consider a  noiseless setting. Another common implementation consists in linearly combining the outputs predicted by the network with the values measured during sampling based on an estimated noise level \cite{eo2018kiki,RN306}. Our final cascade of U-nets model is given by:
\begin{equation}
\widehat{Y}=f_{w_L}[...f_{w_1}[\mathbf{x}_u]], \text{ } \forall w_l \in \{k,i\} \text{ and }  l \in \{1,...,L\} 
\end{equation}
$\widehat{Y}$ is the reconstruction estimated by the model. The loss function used to train the model was simply the mean squared error:

\begin{equation}
    \mathcal{L} = \frac{1}{N}\sum_{i = 1}^{N}||\widehat{Y}^ i - Y^i||^2,
\end{equation}
where $N$ is the number of samples used to compute the loss and the upper script $i$ indicates a sample in this set.

\subsection{Deep Learning Models}
\begin{figure}[!h]
\centering
\includegraphics[width=0.49\textwidth]{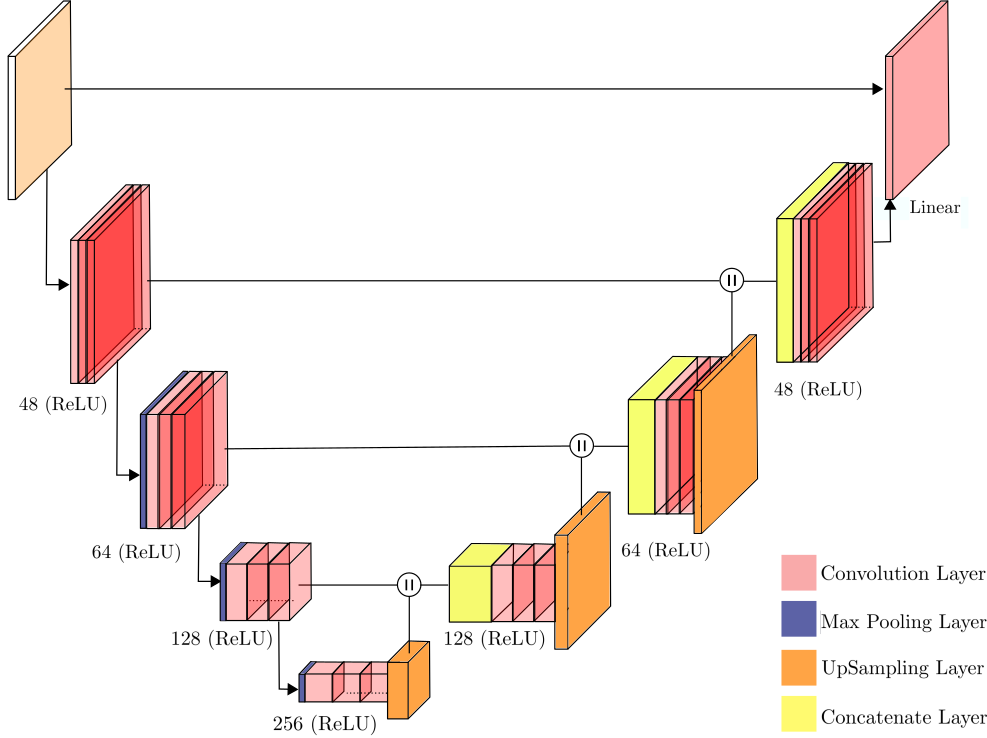}
\caption{The base U-net model architecture. The network receives as input  either single-channel (SC) or multi-channel (MC) k-space data.  This U-net has 22 convolutional layers, three max-pooling layers, three up-sampling layers, and one residual connection. The kernel sizes of the convolutions are $3 \times 3$, with the exception of the final layer, where we use $1 \times 1$ convolutions.}
\label{unet_model}
\end{figure}

Four different models were first investigated in this study. The two-element U-net, termed W-net,  was tested using all four possible domain configurations: a) W-net II, b) W-net KK, c) W-net IK, and d) W-net KI.  The U-net model (Figure \ref{unet_model}) used in this work is a modified version of the originally proposed U-net \cite{RN196} and was designed empirically. Modification was made because, when designing our model, we noticed that a network with less convolutions and convolutional layers yielded similar results compared to more complex models. Our U-net has 22 convolutional layers and 3,000,674 for the SC configuration and 3,011,156 trainable parameters for the MC configuration. The W-net models consist of two cascaded U-nets and thus have twice as many convolutional layers and trainable parameters. 

In the second stage, the best-performing W-net was identified and concatenated with itself to form a four-element WW-net model. We compared WW-net against the four previously described models. The WW-net  model  consists  of  four  cascaded  U-nets  and thus has  four times  as  many  convolutional layers and  trainable parameters as the basic U-net.

In addition, we compared our  W- and WW-net results against the previously published Deep Cascade method \cite{RN306}. We implemented Deep Cascade using six sub-networks and five convolutional layers with $64$ $3\times3$ filters and a final convolutional layer that goes back to the number of channels of the input, \textit{i.e.}, either 2 or 24 depending on the model. Our Deep Cascade implementation had 894,348 parameters for the SC configuration and 978,960 trainable parameters for the MC configuration.  We choose to compare our approaches against Deep Cascade because recent work has demonstrated superior performance when compared (\textit{cf.}, \cite{RN306,souza19a}) to other recently published deep-learning-based MR image reconstruction techniques, such as Dictionary Learning MR Imaging \cite{ravishankar2010mr}, DAGAN \cite{RN307}, KIKI-net \cite{eo2018kiki} and the networks discussed in \cite{RN305,souza2018hybrid}. We used our own implementation of Deep Cascade because the original implementation provided by the authors only worked in the SC configuration.


\subsection{Experimental Setup and Implementation}
Each of the six networks described in the previous subsection were trained four times: once each for the SC and MC configurations and for each of two different acceleration factors, $R$. $R$ is the reciprocal of the fraction of k-space that was sampled. In this work we tested $R=4$ and $R=8$. This resulted in a total of 24 trained models. All models were trained from scratch over 50 epochs using the Adam optimizer \cite{kingma2014adam} with a learning rate of $10^{-3}$ and decay of $10^{-6}$ ~. The networks training were interrupted if the cost function did not improve for five consecutive epochs and some of them completed training prior to 50 epochs, which provided the rationale for our chosen number of epochs. Forty-three volumes (consisting of 11,008 slices) were used for training, eighteen volumes (4,608 slices) for model selection (validation), and 50 volumes (12,800 slices) for testing. A Poisson disc distribution sampling scheme \cite{cook1986stochastic} in the $k_y-k_z$ plane, where the center of k-space was fully sampled within a circle of radius 16  to preserve the low-frequency information, was used. 
The radius of 16 was determined experimentally. During training, the sampling patterns were randomly generated on each epoch for data augmentation purposes. The deep learning reconstructions were compared against the fully-sampled partial Fourier reconstruction reference.  The best SC and MC models were also assessed for a range of acceleration factors extending between $R=2$ and $R=20$.


Our reconstruction models were implemented in Python 3 using the Keras library (\emph{https://keras.io/}) and TensorFlow (\emph{https://www.tensorflow.org/}) as the backend. Training, validation and testing were performed on a seventh generation Intel Core i7 processor with 16 GB of RAM memory and a GTX 1070 graphics processing unit (GPU). The code is publicly available at \emph{https://github.com/rmsouza01/CD-Deep Cascade-MR-Reconstruction}.

\subsection{Performance Metrics and Statistical Analysis}
The reconstructed images were assessed both qualitatively (visual assessment) and quantitatively (performance metrics). Qualitative assessments included a single blinded 
expert (NN) reviewing the resulting images and assessing image artifact. Quantitatively, images were assessed using two commonly used image reconstruction performance metrics: nromalized root mean squared error (NRMSE) and peak signal to noise ratio (pSNR). Also, we used the visual information fidelity (VIF) \cite{sheikh2006image} metric, which was shown to have a strong correlation with radiologist opinion when rating MR image quality \cite{mason2019comparison}.

Lower NRMSE represents better reconstructions, while the opposite is true for pSNR and VIF. Where appropriate, mean $\pm$ standard deviation values were reported. Because the metrics did not follow a normal distribution, statistical significance between the experimental network models was determined using a non-parametric Friedman chi-squared test. Post-hoc testing to assess specific pair-wise differences was performed using a Dunn's test with Bonferroni correction. A $p$-value $< 0.05$ was used as the level of statistical significance.

Processing times of the SC and MC channel configurations were measured across two hundred and fifty-six (256) $218\times170$ image slices using the hardware previously described (see \textit{Section III C}). The average reconstruction time per slice for each of the models was reported.

\section{Results}
A range of the slices towards the edges of the three-dimensional acquisition volumes did not contain anatomy and were basically noise. Although qualitatively agreeing with the noise properties of the reference image (Supplementary Figure 2), reconstruction of these edge slices resulted in large changes in the residual maps and were thus excluded from the quantitative image analysis, leaving a total of $7,500$ slices in the test set. 

The metrics for the SC configuration reconstruction  are summarized in Table \ref{sc_table}. Statistically significant differences were found between the group means ($p<0.01$). Post-hoc testing indicated that Deep Cascade had the overall best metrics for $R=4$, although the differences were small when compared with WW-net IIII. For $R=8$, WW-net IIII obtained the best results. Among the SC configuration, in all experiments, image domain learning methods had superior results in the quantitative analysis, followed by hybrid models and then the k-space only model.   

The performance metrics for the MC configuration reconstruction for $R=4$ and $R=8$ are summarized in Table \ref{mc_table}. Statistically significant differences were observed between group means ($p<0.01$). The post-hoc testing indicated that WW-net IKIK had the overall best metrics for both $R=4$ and $R=8$. Among the MC configuration, in all experiments, hybrid-domain learning methods had superior results in the quantitative analysis. 

\begin{table}[!h]
\caption{\textbf{Single Channel (SC) Configuration:} Average normalized root mean squared error (NRMSE), peak signal to noise ratio (pSNR) and visual information fidelity (VIF) reconstruction results for the SC configuration. Mean $\pm$ standard deviation is reported. The best results for each $R$ factor are emboldened. A Friedman chi-squared test determined statistical significance across the six
experimental models ($p<0.01$) for $R = 8$.
Post-hoc pairwise Dunn's test with Bonferroni correction between the WW-net IIII  and the other five
methods for each $R$ factor was significant for all comparisons ($p<0.01$). Image domain learning methods achieved the best quantitative results.}
\label{sc_table}
\centering
\resizebox{\columnwidth}{!}{%
\begin{tabular}{c|c|c|c}
\hline
 \textbf{Model} &  \textbf{NRMSE} &  \textbf{pSNR (dB)} &  \textbf{VIF}\\ \hline
 \multicolumn{4}{c}{ \textbf{$R=4$}} \\ \hline
\textbf{W-net II}      & $0.0299 \pm 0.0072$ &  $30.6 \pm 1.7$ &  $0.855 \pm 0.071$\\ \hline
\textbf{W-net KK}      & $0.0420 \pm 0.0095$ &  $27.7 \pm 1.8$ &  $0.673 \pm 0.086$\\ \hline
\textbf{W-net IK}      & $0.0324 \pm 0.007$ &  $29.9 \pm 1.6$ &  $0.815 \pm 0.076$\\ \hline
\textbf{W-net KI}    & $0.0348 \pm 0.0083$ &  $29.4 \pm 1.8$ &  $0.783 \pm 0.085$\\ \hline
\textbf{WW-net IIII}   & $0.0287 \pm 0.0070$ &  $31.0 \pm 1.7$ &  $0.889 \pm 0.065$\\ \hline
\textbf{Deep Cascade}  & $\boldsymbol{0.0280 \pm 0.0071}$ &  $\boldsymbol{31.2 \pm 1.7}$ &  $\boldsymbol{0.899 \pm 0.066}$\\ \hline
\multicolumn{4}{c}{ \textbf{$R=8$}} \\ \hline
\textbf{W-net II}     &   $0.0416 \pm 0.0098$ &  $27.8 \pm 1.7$ &  $0.767 \pm 0.083$\\ \hline
\textbf{W-net KK}     &   $0.0528 \pm 0.0107$ &  $25.7 \pm 1.6$ &  $0.582 \pm 0.084$\\ \hline
\textbf{W-net IK}     &   $0.0448 \pm 0.010$ &  $27.1 \pm 1.7$ &  $0.716 \pm 0.090$\\ \hline
\textbf{W-net KI}     &   $0.0456 \pm 0.0105$ &  $27.0 \pm 1.7$ &  $0.703 \pm 0.097$\\ \hline
\textbf{WW-net IIII}  &    $\boldsymbol{0.0405 \pm 0.0097}$ &  $\boldsymbol{28.0 \pm 1.7}$ &  $\boldsymbol{0.795 \pm 0.088}$\\ \hline
\textbf{Deep Cascade} &   $0.0413 \pm 0.0101$ &  $27.9 \pm 1.8$ &  $0.793 \pm 0.084$\\ \hline
\end{tabular}}
\end{table}

\begin{table}[!h]
\caption{\textbf{Multi-Channel (MC) Configuration:} Average normalized root mean squared error (NRMSE), peak signal to noise ratio (pSNR) and visual information fidelity (VIF) reconstruction results for the MC configuration. Mean $\pm$ standard deviation is reported. The best results for each $R$ factor are emboldened. A Friedman chi-squared test determined statistical significance across the six experimental models ($p<0.01$) for both $R$ factors. Post-hoc pairwise Dunn's test with Bonferroni correction between the WW-net IKIK  and the other five methods for each $R$ factor was significant for all comparisons ($p<0.01$). Hybrid learning methods achieved the best quantitative results.\\ }

\label{mc_table}
\centering
\resizebox{\columnwidth}{!}{%
\begin{tabular}{c|c|c|c}
\hline
\textbf{Model} &   \textbf{NRMSE } &  \textbf{pSNR (dB)} &  \textbf{VIF}\\ \hline
\multicolumn{4}{c}{ \textbf{$R=4$}} \\ \hline
\textbf{W-net II}      & $0.0246 \pm 0.0064$ &  $32.3 \pm 1.6$ &  $0.934 \pm 0.048$\\\hline
\textbf{W-net KK}      & $0.0243 \pm 0.0062$ &  $32.5 \pm 1.5$ &  $0.912 \pm 0.048$\\ \hline
\textbf{W-net IK}     & $0.0228 \pm 0.0061$ &  $33.0 \pm 1.6$ &  $0.956 \pm 0.045$\\ \hline
\textbf{W-net KI}      & $0.0230 \pm 0.006$ &  $32.9 \pm 1.5$ &  $0.955 \pm 0.046$\\ \hline
\textbf{WW-net IKIK}   & $\boldsymbol{0.0215 \pm 0.0059}$ &  $\boldsymbol{33.5 \pm 1.6}$ &  $\boldsymbol{0.977 \pm 0.040}$\\ \hline
\textbf{Deep Cascade} & $0.0235 \pm 0.0062$ &  $32.8 \pm 1.6$ &  $0.970 \pm 0.040$\\ \hline
\multicolumn{4}{c}{ \textbf{$R=8$}} \\ \hline
\textbf{W-net II}     & $0.0352 \pm 0.0088$ &  $29.2 \pm 1.7$ &  $0.871 \pm 0.067$\\ \hline
\textbf{W-net KK}     & $0.0380 \pm 0.0089$ &  $28.6 \pm 1.6$ &  $0.784 \pm 0.066$\\ \hline
\textbf{W-net IK}     & $0.0330 \pm 0.0081$ &  $29.8 \pm 1.6$ &  $0.890 \pm 0.062$\\ \hline
\textbf{W-net KI}     & $0.0336 \pm 0.008$ &  $29.6 \pm 1.6$ &  $0.866 \pm 0.068$\\ \hline
\textbf{WW-net IKIK}  & $\boldsymbol{0.0308 \pm 0.0081}$ &  $\boldsymbol{30.4 \pm 1.6}$ &  $\boldsymbol{0.915 \pm 0.064}$\\ \hline
\textbf{Deep Cascade} & $0.0336 \pm 0.0084$ &  $29.6 \pm 1.6$ &  $0.902 \pm 0.059$\\ \hline
\end{tabular}}
\end{table}

Representative sample reconstructed images using the SC and MC configurations for $R=4$ and $R=8$ are depicted in Figures \ref{r4} and \ref{r8}, respectively. Visual assessment of the reconstructed images showed noticeable reconstruction artifacts, particularly with the SC configuration and the W-net KK model. Artifacts are more noticeable at $R=8$.

The arguably best SC model was WW-net IIII and the best MC model was WW-net IKIK. They were trained and tested for a range of acceleration factors ($2 \leq R \leq 20)$. The average NRMSE, pSNR and VIF results are depicted in Figure \ref{metrics_accelerations}. On average the MC WW-net IKIK decreased NRMSE by $29.3\%$ and increased pSNR and VIF by $6.5\%$ and $11.2\%$, respectively, compared to SC 
W-Wnet IIII. Differences were statistically significant ($p<0.01$).
Representative reconstructions for each accleration factor in the SC and MC configurations are depicted in Figures \ref{sc_accelerations} and \ref{mc_accelerations}, respectively.

The average reconstruction time for each of the models assessed are reported in Table \ref{time}. The SC configuration was slower, because it reconstructed each of the 12-channels independently prior to combining them through sum of squares. The slowest model in the SC configuration was WW-net IIII, which took $\approx 450$ ms to reconstruct each slice. The second slowest was Deep Cascade followed by the W-net models. Our MC configuration implementation was not optimal, specially the portion that computes the channel-wise FT (FFT or iFFT), which was implemented through a slow interpreted loop in Python. Although the MC configuration had a sub-optimal implementation, the slowest model required $<65$ ms to reconstruct a slice.

\begin{table}[!h]
\caption{Average reconstruction times for the different models across the single-channel (SC) and multi-channel (MC) configurations. The SC configuration is considerably slower, because it reconstructed each of the 12-channels independently. Note that the reconstruction times roughly double with the depth of the cascade (\textit{i.e.}, W-net \textit{versus} WW-net). For the MC configuration the implementation of the appropriate channel-wise Fourier Transform (direct or inverse) was sub-optimal. Therefore, the processing times did not scale with the cascade depth.}
\label{time}
\centering
\begin{tabular}{c|c|c}
\hline
 \textbf{Model} & \textbf{SC (ms)} & \textbf{MC (ms)} \\ \hline
 \textbf{W-net II} & 222.0 &  39.2 \\ \hline
 \textbf{W-net KK} &  222.0 & 34.4 \\ \hline
\textbf{W-net IK} & 222.0 &  38.8 \\ \hline
\textbf{W-net KI} & 222.0 &   36.4 \\ \hline
\textbf{WW-net IIII/IKIK} & 452.4 & 57.0   \\ \hline
\textbf{Deep Cascade} & 400.8 &  62.4 \\ \hline
\end{tabular}
\end{table}

\section{Discussion}
Our experiments indicated that cascades of U-nets can  improve CS MR reconstruction. In our comparison with the Deep Cascade method that is composed of six flat unrolled sub-networks, our WW-net model (composed of a cascade of four U-nets) achieved statistically significant better results in three out of four experiments (SC, $R=4$ was the exception). In the MC configuration, Deep Cascade was also outperformed by the hybrid W-net models (IK and KI) in terms of pSNR and NRMSE. These results are not surprising, since U-nets are more flexible models that work across different scales when compared to flat convolutional neural networks. Also, a single U-net had been shown to be superior to a flat CNNN model for MR reconstruction when using architectures that produced the same number of feature maps \cite{RN253}. WW-net has $\approx12\times$ more trainable parameters when compared to Deep Cascade. Nevertheless, WW-net and Deep Cascade had similar processing times. Deep Cascade implementation was faster than WW-net in the SC configuration by $\approx 55$~ms per slice reconstructed, while it was slower in the MC configuration by $\approx 5$~ms per slice.

In all of the four experiments, W-net IK models slightly outperformed W-net KI models indicating that it may be advantageous to start the cascade using an image domain network. A similar result using flat unrolled structures was reported in \cite{souza19a}. This finding can be explained by the fact that high frequencies of k-space are less densely sampled, potentially resulting in regions where the convolutional kernel would have no signal to operate upon. By starting with an image domain CNN block and because of the global property of the FT, the output of this network has a corresponding k-space that is now complete, effectively mitigating the problem of regions with no samples for the convolution kernels to operate on. 

\begin{figure*}[!h]
\centering
\includegraphics[width=1.0\textwidth]{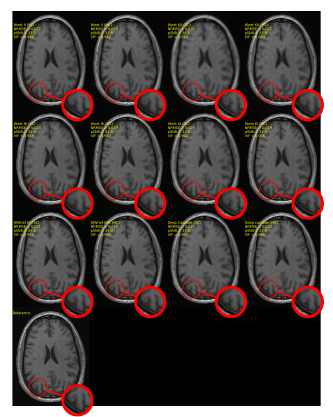}
\caption{Illustrative example of reconstructed images using the same acceleration rate ($R=4$) on a single-channel (SC) or multi-channel (MC) configurations. Blurred images and poor delineations between lateral ventricle and white matter boundaries were detected only on SC W-net KK. Apart from SC W-net KK, most reconstructed images were in agreement with the reference image (VIF > 0.9). Overall best reconstructed images were obtained with MC WW-net IKIK, while Deep Cascade demonstrated the best result among models trained with the SC configuration. Zoomed-in tools were used to magnify a selected part of the brain revealing observable differences in image quality between the six different reconstruction methods. }
\label{r4}
\end{figure*}

\begin{figure*}[!h]
\centering
\includegraphics[width=1.0\textwidth]{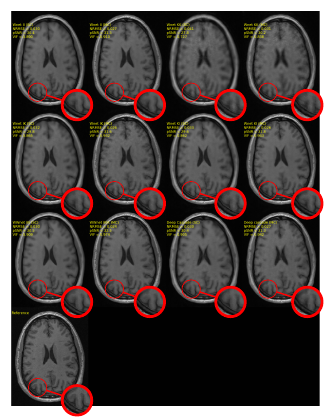}
\caption{ Illustrative example of reconstructed images using the same acceleration rate ($R=8$) on a single-channel (SC) or multi-channel (MC) configurations. For both configurations, W-net KK reconstructed images had poor delineations between brain structures. Note visible differences in anatomical delineation between lateral ventricles and white matter. Results from qualitative and our expert visual assessment, showed that reconstructed images with MC WW-net IKIK had the best overall quality. Zoomed-in tools were used to magnify a selected part of the brain image revealing observable differences in image quality between six different reconstruction methods. }
\label{r8}
\end{figure*}

The SC configuration results indicated that image-learning methods are better suited for the reconstruction of these kind of data followed by hybrid learning approaches and then sensor (k-space) learning models. W-net KK had clear blurring and ringing artifacts that made it especially difficult to distinguish the transition between white-matter and gray-matter tissue (Figures \ref{r4} and \ref{r8}). The ranking of techniques for the MC configuration was different. Hybrid learning models achieved the best quantitative metrics. The observed performance boost of hybrid and sensor (k-space) learning methods can be attributed to the fact that now the k-space  (sub-)networks learn not just potential correlations within the same k-space, but also correlations present across coil channels. Correlations in space across channel are known to be strong and are the underlying basis of PI techniques \cite{grappa}.

The results of the MC configuration experiments were superior to the SC configuration results (Tables \ref{sc_table} and \ref{mc_table}). The best MC configuration metrics  reduced NRMSE by $29.3\%$ and increased pSNR and VIF by $6.5\%$ and $11.2\%$, respectively, compared to the best SC model. This observation is explained by the fact that the MC configuration looks at all channels simultaneously. The advantage of the SC configuration is its flexibility. A properly trained SC network can work for an arbitrary number of coil channels (see Supplementary Figure 3 for an example using a 32-channel coil). In contrast, for the MC configuration, it is necessary to train one model for every coil-channel configuration. The models trained using the MC configuration were between $4 \times$ and $8 \times$ faster than the SC configuration models and that difference is expected to increase when using larger number of channels. Nonetheless, by using modern GPUs and by optimizing the reconstruction code, both models have the potential for online, \textit{i.e.}, while patient is still in the scanner, MR reconstruction.  

\begin{figure*}[!h]
\centering
\includegraphics[width=1.0\textwidth]{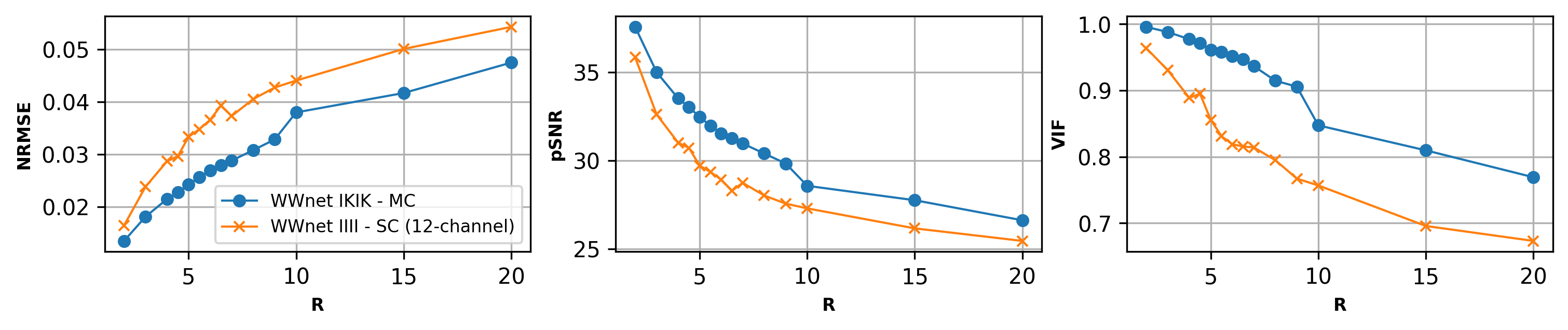}
\caption{Average normalized root mean squared error (NRMSE), peak signal to noise ratio (pSNR), and visual information fidelity (VIF) for the SC WW-net IIII model (orange curve) and the MC WW-net IKIK model (blue curve) computed across a range of acceleration factors. Differences were statistically significant ($p<0.01$).}
\label{metrics_accelerations}
\end{figure*}

\begin{figure*}[!h]
\centering
\includegraphics[width=1.0\textwidth]{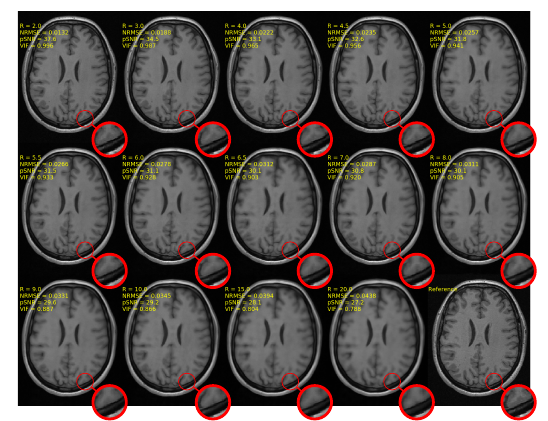}
\caption{Single-channel (SC) WW-net IIII model can demonstrate detectable differences in image quality across fourteen different acceleration rates $R = {2, 3, 4, 4.5, 5, 5.5, 6, 6.5, 7, 8, 9, 10, 15, 20}$. Higher acceleration rates ($R>8$) resulted in progressive deterioration in the quality of reconstructed image as white/gray-matter delineation (see zoomed-in areas).}
\label{sc_accelerations}
\end{figure*}

\begin{figure*}[!h]
\centering
\includegraphics[width=1.0\textwidth]{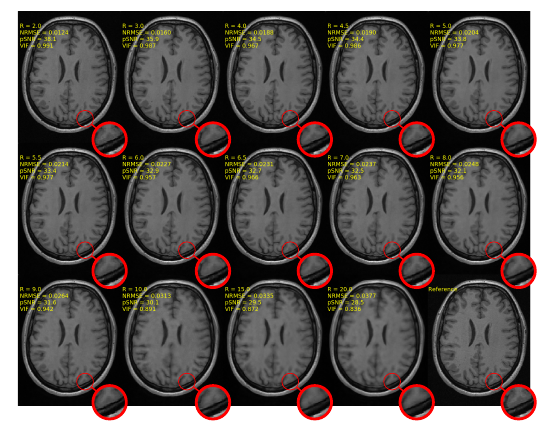}
\caption{Reconstructed images from multi-channel (MC) WW-net IKIK model can also vary in quality across fourteen acceleration rates $R = {2, 3, 4, 4.5, 5, 5.5, 6, 6.5, 7, 8, 9, 10, 15, 20}$. Similarly to the single-channel example (Figure \ref{sc_accelerations}, higher acceleration rates ($R>10$) resulted in progressively lower image quality. Note the marginal lines between white and gray-matter become severely blurred at higher acceleration rates (see zoomed-in areas).}
\label{mc_accelerations}
\end{figure*}

Visual inspection of the images agree with the quantitative results. Visual differences between SC and MC reconstructions are more noticeable at higher acceleration factors (Figure \ref{r8}). The VIF metric, which is correlated with the radiologist assessment of image quality \cite{mason2019comparison}, has a larger difference between SC and MC configurations when compared at $R > 5$ (Figure \ref{metrics_accelerations}). Assuming that an image with VIF $> 0.9$ is good enough to be incorporated into the clinical setting, the SC configuration would allow acceleration factors of up to $R=4.5$, while the MC configuration would allow accelerations of up to $R=9$ (Figure \ref{metrics_accelerations}) when using a 12-channel coil. The usage of more sophisticated coils with more elements could potentially allow for further acceleration.

\section{Conclusions}
In this work, we investigated cascades of U-nets across different domain configurations for MR reconstruction. Two different configurations were investigated: the SC and the MC configurations. Our results indicate that image domain learning approaches are advantageous when processing channels independently (SC configuration), while hybrid approaches are better when reconstructing all channels simultaneously (MC configuration).  The MC configuration also proved to be considerably faster than the SC configuration with an speed difference proportional to the number of coil channels. The SC configuration, however, is more flexible than the MC configuration, because it is independent of the number of coil channels. Our WW-net (IIII for SC and IKIK for MC) method outperformed the state-of-the-art Deep Cascade method in three of the four comparisons. Unlike previous studies (\textit{cf}. \cite{souza2018hybrid,eo2018kiki}), our investigations indicated that starting the cascade of U-nets with an image domain network for MC data leads to better results. Future studies should investigate the SC and MC configurations using a range of different coils (\textit{e.g.}, 4-channel, 8-channel, \textit{etc}.). Also, the optimal domain configuration for the sub-networks that compose the cascade of U-nets, which is a problem that grows exponentially ($2^M$ where $M$ is the number of sub-networks) is not yet known. 





\bibliographystyle{IEEEtran}
\bibliography{sample}








\end{document}